\documentclass[a4paper]{jpconf}
\usepackage{graphicx}
\begin{document}
\title{Jet structure modifications in heavy-ion collisions with JEWEL}

\author{Raghav Kunnawalkam Elayavalli}

\address{Department of Physics and Astronomy, Rutgers University, Piscataway, New Jersey 08854, USA}

\ead{raghav.k.e@cern.ch}

\begin{abstract}
	Key features of jet-medium interactions in heavy-ion collisions are modifications to the jet structure. Recent results from experiments at the LHC and RHIC have motivated several theoretical calculations and monte carlo models towards predicting these observables simultaneously. In this report, the recoil picture in \textsc{Jewel} is summarized and two independent procedures through which background subtraction can be performed in \textsc{Jewel} are introduced. Information of the medium recoil in \textsc{Jewel} significantly improves its description of several jet shape measurements. 
\end{abstract}

\section{Introduction}
	The qualitative effect of jet quenching through the measurements of the jet nuclear modification factor ($R_{\rm{AA}}$) are confirmed with Run1 data at the LHC~\cite{atlasraa, aliceraa, cmsraa}. While the $R_{\rm{AA}}$ shows a clear and expected trend in the medium induced effects from central to peripheral events, the energy loss on a jet-by-jet level is not characterized. This motivated several measurements that probed the inner-structure of jets such as jet shapes~\cite{cmsjetshapes1, alicepreljetshapes, cmsjetshapes2}, searches for the quenched energy away from the jet axis~\cite{cmsmissingpt} and fragmentation functions~\cite{cmsfragfunc, atlasfragfunc} to name a few. From these detailed measurements, jets thats propagate the quark gluon plasma (QGP) were perceived as getting broader, losing energy inside the jet cone, increased multiplicity of low transverse momenta ($p_{T}$) particles around the periphery of the jet and several other consistent observations when compared to jets in pp collisions. The subjet groomed momentum fraction was recently measured as a function of the jet $p_{T}$ and event centrality at CMS~\cite{cmssplittingfunction}, concluding that jets were more asymmetrically split in heavy-ion collisions as opposed to those in pp collisions.  
	
	All these aforementioned results point to quantitative features of jet quenching, hence it is imperative that any phenomenological model describing the physics of QGP has to predict the general behavior and trend of these observables. Currently there are several theoretical calculations and monte carlo models available on the market (see~\cite{yacine_review} for a comprehensive review). These results offer a unique way of discriminating between these models.
	
	\textsc{Jewel}~\cite{jewel} is a monte carlo framework utilizing a perturbative quantum chromodynamical (pQCD) implementation of in-medium energy loss. 
	The latest version of \textsc{Jewel} is interfaced with \textsc{Pythia6}~\cite{pythia6} and the full framework operates as follows: 
	\begin{itemize}
		\item \textsc{Pythia} produces the di-jet hard scattering and initial state radiation 
		\item \textsc{Jewel} takes these hard scattered partons and proceeds with the final state shower
		\item \textsc{Pythia} takes over for hadronization and hadron decays 
		\item These events are then analyzed with the \textsc{Rivet}~\cite{rivet} analysis framework 
	\end{itemize}  

	 The medium in \textsc{Jewel} is characterized as a thermal distribution of scattering centers. The hard scattered partons undergo microscopic interactions that are estimated with perturbative matrix elements and the aforementioned parton shower modifications, giving rise to elastic and inelastic energy losses.  At each interaction, a recoiled parton is created and in the current version of \textsc{Jewel}, propagates without any further interaction. This is a limiting case, since in reality these recoiled partons can further interact with the medium. Detailed descriptions of the monte carlo implementation in \textsc{Jewel} and corresponding studies are available here~\cite{jewel,jewelphysics}.
	
	The recoiled parton's energy is comprised of both the collisional part due from the hard scattered parton, and the thermal component of the scattering centers. This extraneous energy gets clustered in along with the other particles in the event and becomes part of the jets. Hence, when comparing data with \textsc{Jewel} (including recoils), a slight mismatch appears for inter-jet observables due to the experimental implementation of the background subtraction. Since \textsc{Jewel} does not simulate a full heavy ion event, the exact method utilized by different experiments cannot be reliably implemented. However, due to the microscopic nature of the interactions, the exact amount of background energy and momentum is easily estimated as the thermal component of the scattering centers before interaction. Any such subtraction techniques are only viable for infrared safe observables since the energy corresponding to the scattering centers are before hadronization effects. 
	
\section{Background subtraction in \textsc{Jewel}}
	The information of the scattering centers before the interaction are first included in the event record with a separate tag, so as to not disturb the jet clustering. Simultaneously dummy particles with very small momenta and positions corresponding to the scattering centers are introduced to the final state particle collection. With this information, one can associate the scattering centers involved with the corresponding jet and hence proceed with background subtraction with either of the following methods: 
	\begin{itemize}
		\item 4MomSub: Once the jet constituents are matched in position to their corresponding scattering centers, a simple four-momenta vectorial subtraction is performed to remove the background contribution to the respective jet. 
		\item GridSub: A finite resolution grid is superimposed on the event confirning the jet constituents and their scattering centers in grid cells. Inside each cell, the momenta of the scattering centers are vectorially subtracted and the jet is finally clustered with each cell as an input pseudo-particle. For the case when cells only contain scattering centers, their momentum is set to zero before the clustering procedure.  
	\end{itemize} 

	The 4MomSub is recommended when possible since it is closer to a true background subtraction. The GridSub method is employed in cases when observables require the information of jet constituents, and is very similar to experiments where the results are constrained by the detector's finite resolution.  Detailed studies of the systematic uncertainties introduced by the background subtraction procedures are in preparation.  

\section{Comparisons with data}
	\begin{figure}[h] 
		\centering
		\includegraphics[width=0.6\textwidth]{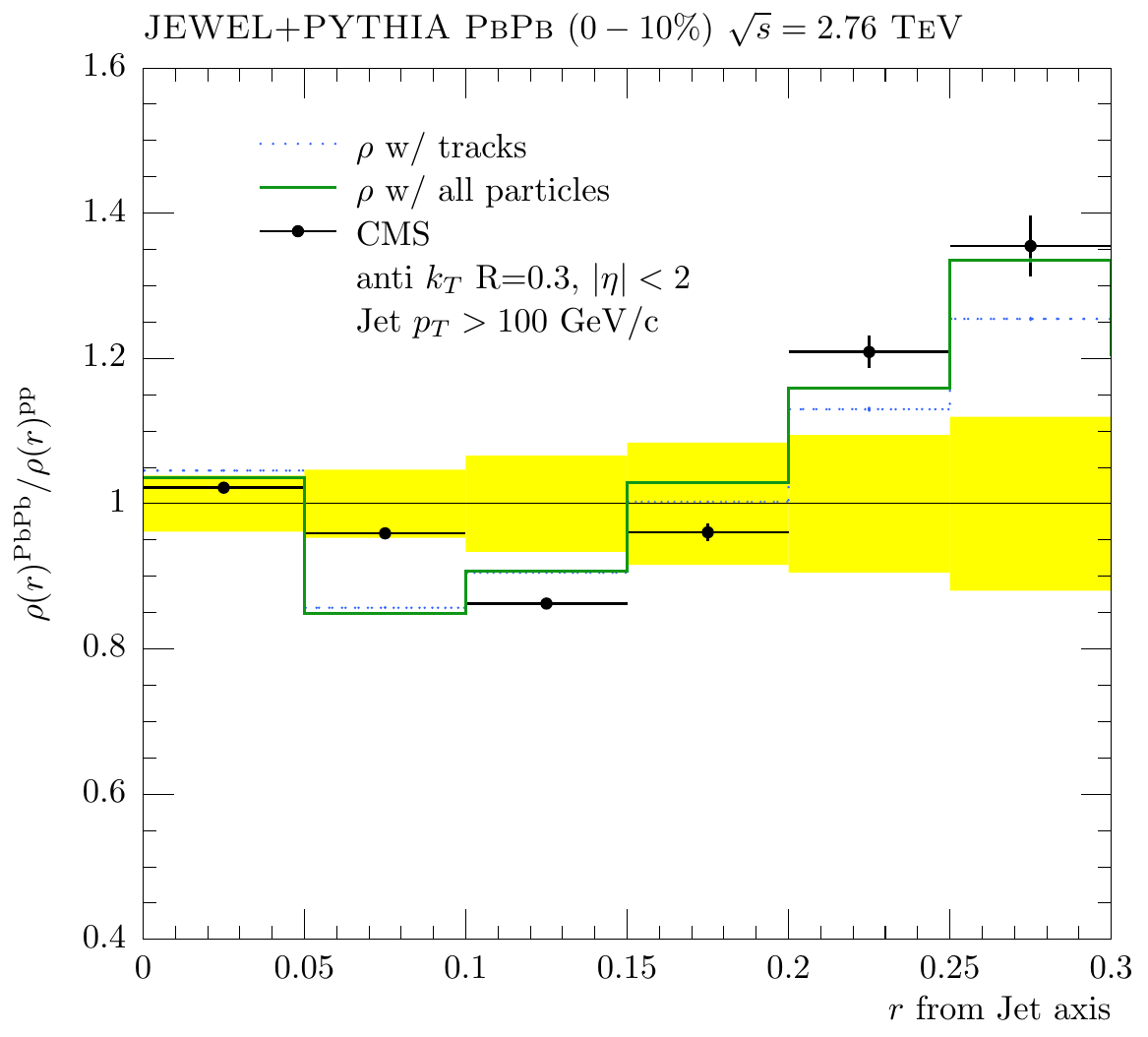} 
		\caption{Differential jet shape ratio of PbPb to pp collisions comparing \textsc{Jewel+Pythia} predictions  with CMS data (black points). For the monte carlo predictions, the density estimated with only tracks vs. using all final state particles is shown in the blue dotted and green solid lines respectively. The Data systematic uncertainties are shown in the yellow boxes around unity.}
		\label{fig:cmsJetShape}
	\end{figure}
	
	The ratio of the track yield in annuli around the jet axis in most central PbPb collisions at 2.76 TeV compared to pp collisions are shown in Fig.~\ref{fig:cmsJetShape}. The CMS data points~\cite{cmsjetshapes1} are shown in black markers while \textsc{Jewel}+\textsc{Pythia} predictions are presented in blue dotted lines where the density is estimated with tracks and with green solid lines for density estimated with all final state particles. Due to the partonic nature of the background in \textsc{Jewel}, the subtraction technique is efficient when considering  both neutral and charged in the event. The data systematic uncertainty is presented in the yellow shaded region. The general trend of the data is reproduced by \textsc{Jewel}+\textsc{Pythia} after subtraction using the 4MomSub method for anti-k$_{t} R = 0.3$ jets clustered with the \textsc{Fastjet}~\cite{fastjet} toolkit. As expected the agreement gets better with the data at large $r$ when the density is estimated with all final state particles.
 			
	\begin{figure}[h] 
		\centering
		\includegraphics[width=0.47\textwidth]{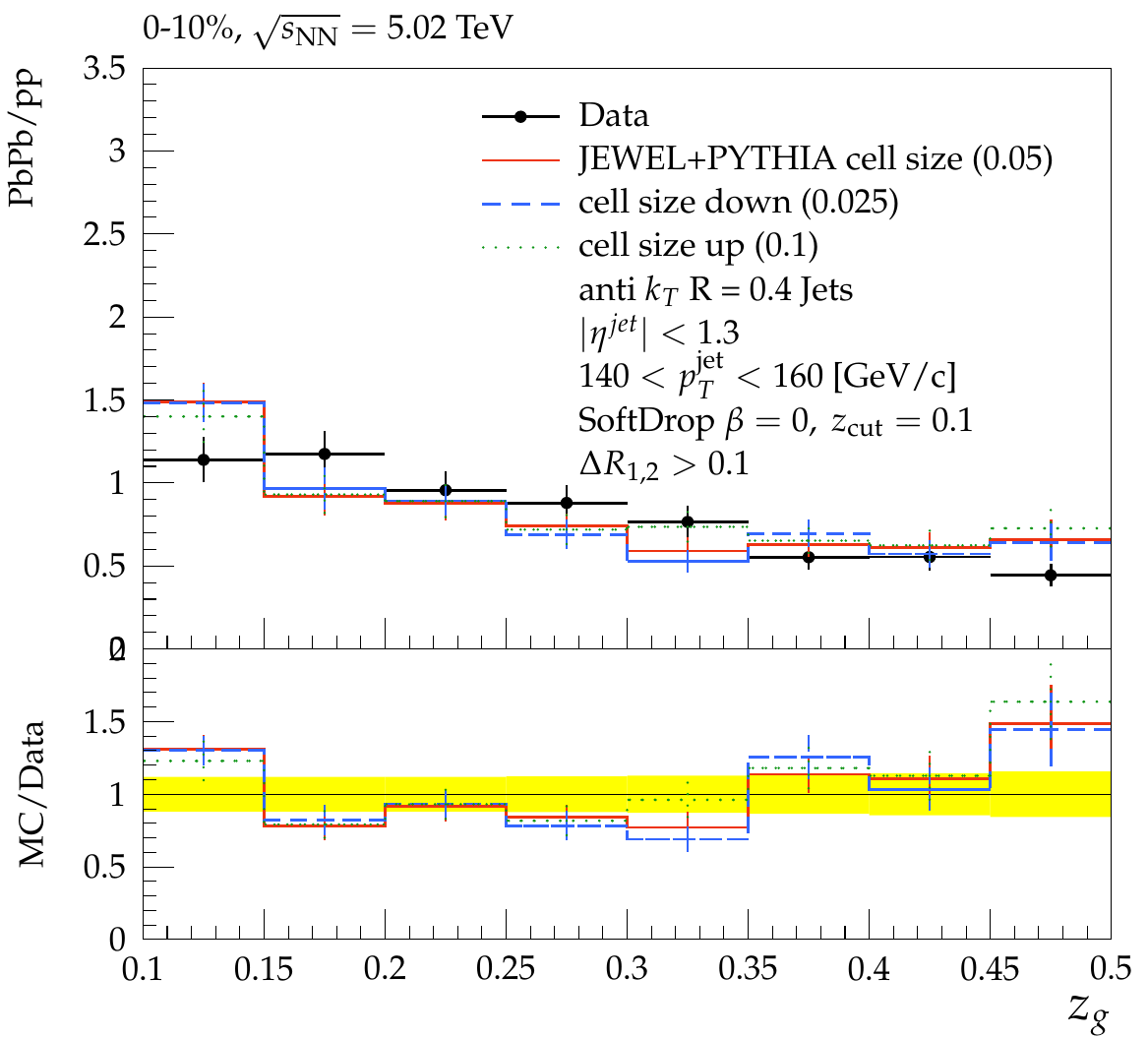} 
		\includegraphics[width=0.47\textwidth]{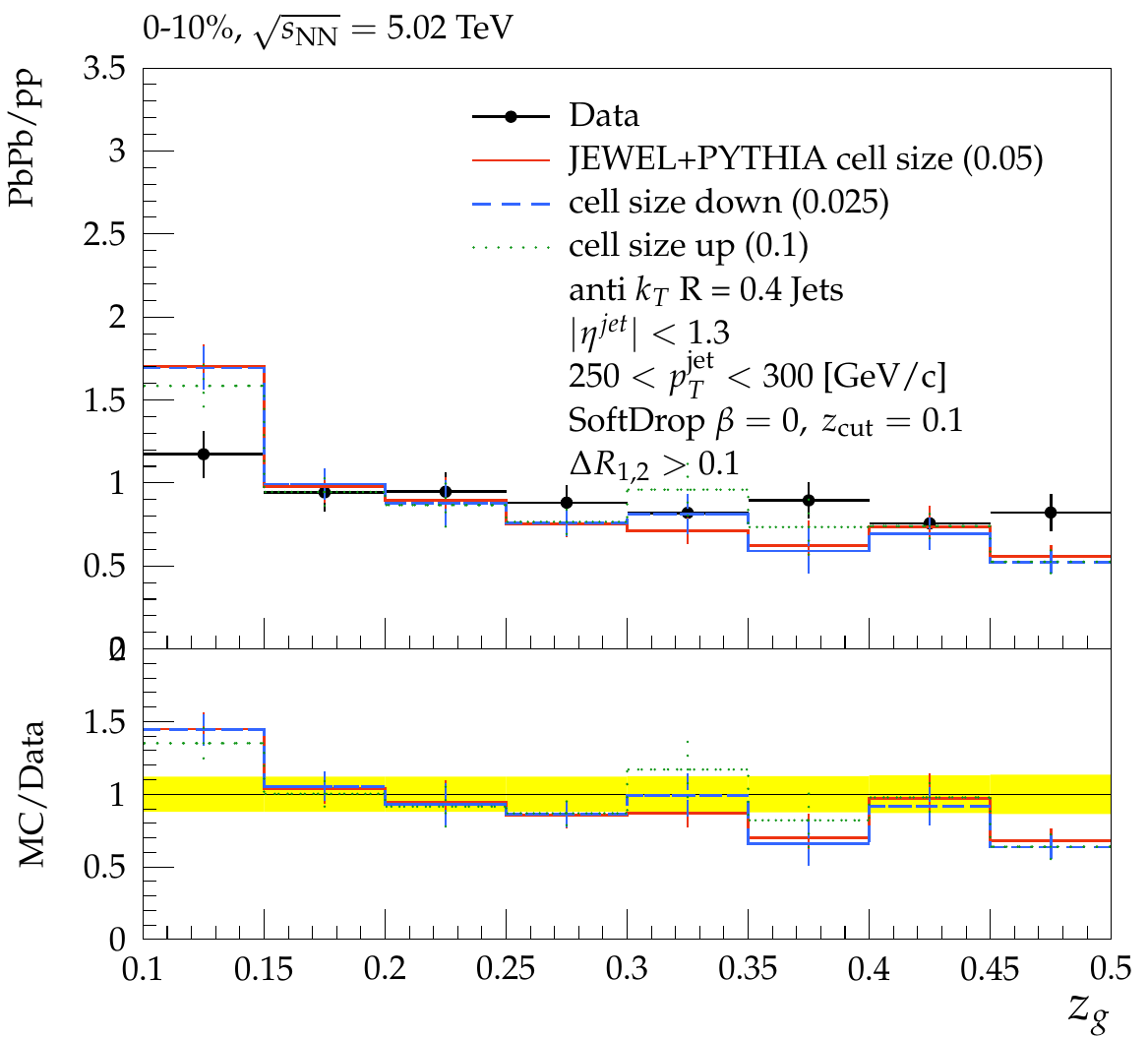} 
		\caption{Comparison of \textsc{Jewel+Pythia} predictions (colored lines) with CMS data (black markers) for the ratio of the subjet groomed momentum fraction distributions in central PbPb to pp events. The low and high $p_{T}$ ranges are shown on the left and right respectively. The bottom panels presents the ratio of the monte carlo predictions with data.}
		\label{fig:cmsSplitting}
	\end{figure}	

	The \textsc{Jewel}+\textsc{Pythia} predictions for the ratios of the subjet groomed momentum fractions in PbPb to pp collisions are compared with CMS data~\cite{cmssplittingfunction} as shown in Fig.~\ref{fig:cmsSplitting}. The subjet groomed momentum fractions are estimated via the softdrop framework~\cite{softdrop}, for low (left) and high (right) $p_{T}$ anti-k$_{t} \ R = 0.4$ jets respectively in Fig.~\ref{fig:cmsSplitting}. The ratio of \textsc{Jewel}+\textsc{Pythia} to data is shown in the bottom panels where the yellow shaded region represents the total uncertainty in the data points. The systematics in the \textsc{Jewel}+\textsc{Pythia} predictions are shown by varying the grid resolution by a factor of two.  \textsc{Jewel}+\textsc{Pythia} also reproduces the general trend corresponding to more asymmetric jet splittings in PbPb jets at low $p_{T}$ and more symmetric splittings as the jet $p_{T}$ increases.           

\section{Conclusions}
	Including the recoils in \textsc{Jewel} along with the background subtraction procedures allows the recovery of the jet energy   acquired by the medium and its response to jets. This interplay makes \textsc{Jewel} capable of reproducing the qualitative trend in data related to jet structure modifications. This new class of jet observables that probe the medium-jet interaction highlights the next era in jet tomography  in heavy ion collisions. 

	This work was done in collaboration with Dr. Korinna Christine Zapp. RKE thanks the CERN theory department for its hospitality. This work was supported by Funda\c{c}\~{a}o para a Ci\^{e}ncia e a Tecnologia (Portugal) under project  CERN/FIS-NUC/0049/2015, postdoctoral fellowship SFRH/BPD/102844/2014 (KCZ) and by the European Union as part of the FP7 Marie Curie Initial Training Network MCnetITN (PITN-GA-2012-315877) (RKE). RKE also acknowledges support from the National Science Foundation under Grant No. 1352081, Claude Lovelace fellowship at Rutgers University and thanks Dr. Chun Shen for providing the initial hydrodynamics parameters for the event generation at 5.02 TeV.

\section*{References}
\bibliography{jewel_raghav_hq2016}

\end{document}